\documentclass[12pt]{article}
\begin{document}
\title{Matters of Gravity:Dark Matter and Cosmological Constant}
\author{B.G. Sidharth\\
International Institute for Applicable Mathematics \& Information Sciences\\
Hyderabad (India) \& Udine (Italy)\\
B.M. Birla Science Centre, Adarsh Nagar, Hyderabad - 500 063
(India)}
\date{}
\maketitle
\begin{abstract}
We consider two problems that have vexed physicists for several
decades -- dark matter and the cosmological constant. The problem
has been that the former has not been detected while the latter
gives a far higher value than detected by observation. We argue that
a time varying gravitational constant obviates the former problem,
while the latter problem can be circumvented by considering the
average density of the universe.
\end{abstract}
\vspace{5 mm}
\begin{flushleft}
98.80H; 04.40.-b
\end{flushleft}
\section{Introduction}
Two of the problems that have been plaguing gravitation theory for a
long time are those of Dark Matter and the Cosmological Constant.
Let us see what these problems are:\\
It is well known that F. Zwicky introduced the concept of dark
matter to account for the anomalous rotation curves of the galaxies
\cite{narlikarcos,tduniv}. The problem was that according to the
usual Newtonian Dynamics the velocities of the stars at the edges of
galaxies should fall with distance as in Keplarian orbits, roughly
according to
\begin{equation}
v \approx \sqrt{\frac{GM}{r}}\label{e1}
\end{equation}
where $M$ is the mass of the galaxy, $r$ the distance from the
centre of the galaxy of the outlying star and $v$ the tangential
velocity of the star. Observations however indicated that the
velocity curves flatten out, rather than follow the law (\ref{e1}).
This necessitated the introduction of the concept of dark matter
which would take care of the discrepancy without modifying Newtonian
dynamics. However even after nearly eight decades, dark matter has
not been detected, even though there have been any number of
candidates proposed for this, for example SUSY particles, massive
neutrinos, undetectable brown dwarf stars,
even black holes and so on.\\
Very recent developments are even more startling. These concern the
rotating dwarf galaxies, which are satellites of the Milky Way
\cite{metz1,metz2}. These studies throw up a big puzzle. On the one
hand these dwarf satellites cannot contain any dark matter and on
the other hand the stars in the satellite galaxies are observed to
be moving much faster than predicted by Newtonian dynamics, exactly
as in the case of the galaxies themselves. Metz, Kroupa, Theis,
Hensler and Jerjen conclude that the only explanation lies in
rejecting dark matter and Newtonian gravitation. Indeed a well known
Astrophysicist, R. Sanders from the University of Groningen
commenting on these studies notes \cite{physorg}, "The authors of
this paper make a strong argument. Their result is entirely
consistent with the expectations of modified Newtonian dynamics
(MOND), but completely opposite to the predictions of the dark
matter hypothesis. Rarely is
an observational test so definite."\\
A further evidence has very recently come to light due to an
observation of star light at the fringes of the galaxy by Petrosian
and others. This too goes against Dark Matter \cite{eric}. One of
the arguments which explain the observations, but from what has been
called the MOND point of view has been put forward by Milgrom.
According to this hypotheses, a test particle at a distance $r$ from
a large mass $M$ is subject to the acceleration $a$ given by
\begin{equation}
a^2/a_0 = MGr^{-2},\label{3em1}
\end{equation}
where $a_0$ is an acceleration such that standard Newtonian dynamics
is a good approximation only for accelerations much larger than
$a_0$. The above equation however would be true when $a$ is much
less than $a_0$. Both the statements in (\ref{3em1}) can be combined
in the heuristic relation
\begin{equation}
\mu (a/a_0) a = MGr^{-2}\label{3em2}
\end{equation}
In (\ref{3em2}) $\mu(x) \approx 1$ when $x >> 1, \, \mbox{and}\,
\mu(x) \approx x$ when $x << 1$. It must be stressed that
(\ref{3em1}) or (\ref{3em2}) are not deduced from any theory, but
rather are an ad hoc prescription to explain observations.
Interestingly it must be mentioned that most of the implications of
Modified Newtonian Dynamics or MOND do not
depend strongly on the exact form of $\mu$.\\
It can then be shown that the problem of galactic velocities is now
solved \cite{mil1,mil2,mil3,mil4,mil5}. Nevertheless, most
physicists are not comfortable with MOND because of the ad hoc
nature of (\ref{3em1}) and (\ref{3em2}).
\section{Varying $G$ Dynamics}
We now come to the cosmological model described by the author in
1997 (Cf.ref.\cite{ijmpa1998,tduniv} and several references
therein), in which the universe, under the influence of dark energy
would be accelerating with a small acceleration. Several other
astrophysical relations, some of them hitherto inexplicable such as
the Weinberg formula giving the pion mass in terms of the Hubble
constant were also deduced in this model (Cf.also ref.\cite{cu} and
references therein). While all this was exactly opposite to the then
established theory, it is well known that the picture was
observationally confirmed soon thereafter through the work of
Perlmutter and others (Cf.ref.\cite{cu}). Interestingly, in this
model Newton's
gravitational constant varied inversely with time.\\
Cosmologies with time varying $G$ have been considered in the past,
for example in the Brans-Dicke theory or in the Dirac large number
theory or by Hoyle \cite{barrowparsons,narfpl,narburbridge,5,6}. In
the case of the Dirac cosmology, the motivation was Dirac's
observation that the supposedly large number coincidences involving
$N \sim 10^{80}$, the number of elementary particles in the universe
had an underlying message if it is recognized that
\begin{equation}
\sqrt{N} \propto T\label{3ea1}
\end{equation}
where $T$ is the age of the universe. Equation (\ref{3ea1}) too
leads to a $G$ decreasing inversely
with time as we will now show. We follow a route slightly different from that of Dirac.\\
From (\ref{3ea1}) it can easily be seen that
\begin{equation}
T = \sqrt{N} \tau\label{5}
\end{equation}
where $\tau$ is a typical Compton time of an elementary particle
$\sim 10^{-23}secs$, because $T$, the present age of the universe is
$\sim 10^{17}secs$. We also use the following relation for a
uniformly expanding Friedman Universe
\begin{equation}
\dot{R}^2 = \frac{8 \pi}{3} G \, R^2 \rho\label{6}
\end{equation}
where $R$ is the radius of the universe and $\rho$ its density. We
remember that
\begin{equation}
\rho = \frac{3M}{4 \pi R^3} \, \mbox{and} \, M = Nm\label{7}
\end{equation}
where $M$ is the mass of the universe, and $m$ is the mass of an
elementary particle $\sim 10^{-25}gm$
(Cf.ref.\cite{weinberggravcos}).\\
Use of (\ref{7}) in (\ref{6}) leads to another well known relation
\cite{ruffini}
\begin{equation}
R = \frac{GM}{c^2}\label{8}
\end{equation}
because $\dot{R} = c$. Further dividing both sides of (\ref{5}) by
$c$ we get the famous Weyl-Eddington relation
\begin{equation}
R = \sqrt{N} l\label{9}
\end{equation}
where $l = \tau /c$ is a typical Compton length $\sim 10^{-13}cms$.\\
Use of (\ref{7}) and (\ref{9}) in (\ref{8}) now leads to
\begin{equation}
G = \frac{c^2 l}{\sqrt{N}m} = \left(\frac{c^2 l \tau}{m}\right)
\cdot \frac{1}{T} \equiv \frac{G_0}{T}\label{10}
\end{equation}
Equation (\ref{10}) gives the above stated inverse dependence of the
gravitational constant $G$ on time, which Dirac obtained. On the
other hand this same relation was obtained by a different route in
the author's dark energy -- fluctuations cosmology in 1997. This
work, particularly in the context of the Planck scale has been there
for many years in the literature (Cf.\cite{cu,tduniv,uof} and
references therein). Suffice to say that all the supposedly so
called accidental Large Number Relations like (\ref{9}) as also the
inexplicable Weinberg formula which relates the Hubble constant to
the mass of a pion, follow as deductions in this
cosmology. The above references give a comprehensive picture.\\
The Brans-Dicke cosmology arose from the work of Jordan who was
motivated by Dirac's ideas to try and modify General Relativity
suitably. In this scheme the variation of $G$ could be obtained from
a scalar field $\phi$ which would satisfy a conservation law. This
scalar tensor gravity theory was further developed by Brans and
Dicke, in which $G$ was inversely proportional to the variable field
$\phi$. (It may be mentioned
that more recently the ideas of Brans and Dicke have been further generalized.)\\
In the Hoyle-Narlikar steady state model, it was assumed that in the
Machian sense the inertia of a particle originates from the rest of
the matter present in the universe. This again leads to a variable
$G$. The above references give further details of these various
schemes and their shortcomings which have
lead to their falling out of favour.\\
In any case, our starting point is, equation (\ref{10}) where $T$ is
time (the age of the universe) and $G_0$ is a constant. Furthermore,
other routine effects like the precession of the perihelion of
Mercury and the bending of light and so on have also explained with
(\ref{10}) and furthermore there is observational evidence for
(\ref{10}) (Cf. \cite{tduniv,cu,uzan,bgsnc115b,bgsfpl}); that
described various observational evidences for the variation of $G$,
for example from solar system observations, from cosmological
observations and even from the
palaeontological studies point of view).\\
With this background, we now mention some further tests for equation
(\ref{10}).\\
This could explain the other General Relativistic effects like the
shortening of the period of binary pulsars and so on
(Cf.ref.\cite{cu,tduniv,bgsnc115b,bgsfpl} and other references
therein). Moreover, we could now also explain, the otherwise
inexplicable anomalous acceleration of the Pioneer space crafts
(Cf.ref.\cite{tduniv} for details). We will briefly revisit some of these effects later.\\
We now come to the problem of galactic rotational curves mentioned
earlier (cf.ref.\cite{narlikarcos}). We would expect, on the basis
of straightforward dynamics that the rotational velocities at the
edges of galaxies would fall off according to
\begin{equation}
v^2 \approx \frac{GM}{r}\label{3ey33}
\end{equation}
which is (\ref{e1}). However it is found that the velocities tend to
a constant value,
\begin{equation}
v \sim 300km/sec\label{3ey34}
\end{equation}
This, as noted, has lead to the postulation of the as yet undetected
additional matter alluded to, the so called dark matter.(However for
an alternative view point Cf.\cite{sivaramfpl93}).\\
In any case let us now consider (\ref{10}) in the context of the
usual Keplarian orbit \cite{gold}:
\begin{equation}
\frac{1}{r} = \frac{GMm^2}{l^2}\label{A}
\end{equation}
Let us now differentiate (\ref{A}) keeping in mind Equation
(\ref{10}). This gives us
\begin{equation}
\dot{r} = \frac{G}{t_0} \left(\frac{Mm}{l^2}\right) r^2 =
\frac{r}{t_0}\label{B}
\end{equation}
From (\ref{B}) we get
\begin{equation}
\ddot{r} = \frac{\dot{r}}{t_0} = \frac{r}{t^2_0}\label{C}
\end{equation}
The point is that we recover the usual Newtonian Dynamics with a
constant $G$ if $t_0$ becoming infinite in (\ref{B}) or (\ref{C}).
If we use (\ref{C}), we will get, as can be easily checked
\begin{equation}
v \approx \left(\frac{r^2}{t^2_o} + \frac{GM}{r}\right)^{1/2}
\label{3ey37}
\end{equation}
So (\ref{3ey37}) replaces (\ref{e1}) in this model. This shows that
as long as
\begin{equation}
\frac{r^2}{t_0^2} < < \frac{GM}{r},\label{20}
\end{equation}
Newtonian dynamics holds. But when the first term on the left side
of (\ref{20}) becomes of the order of the second (or greater), the
new dynamical effects come in.\\
For example from (\ref{3ey37}) it is easily seen that at distances
well within the edge of a typical galaxy, that is $r < 10^{23}cms$
the usual equation (\ref{3ey33}) holds but as we reach the edge and
beyond, that is for $r \geq 10^{24}cms$ we have $v \sim 10^7 cms$
per second, in agreement with (\ref{3ey34}). In fact as can be seen
from (\ref{3ey37}), the first term in the square root has an extra
contribution (due to the varying $G$) which exceeds the second term
as we approach the galactic edge, as if there is an extra mass, that
much more.\\
We can estimate this "effective" mass, $M'$ say, as follows: We have
from (\ref{3ey37}),
$$\frac{GM'}{r} = v^2 \approx \frac{r^2}{t^2_0} + \frac{GM}{r}$$
Whence $\Delta M = M' - M$ is given by
$$G\Delta M = r^3 / t^2_0$$
We can easily calculate that this gives for $r \leq 10^{24}cm$, at
the outer edge of the galaxy,
$$\Delta M \geq 10 M$$
in agreement with estimates.\\
We would like to stress that the same conclusions will apply to the
latest observations of the satellite galaxies (without requiring any
dark matter). Let us for example consider the Megallanic clouds
\cite{stave}. In this case, as we approach their edges, the first
term within the square root on the right side of (\ref{3ey37}) or
the left term of (\ref{20}) already becomes of the order of the
second term, leading to the new non Newtonian effects.\\
{\bf A remark:} Equation (\ref{C}) at the scale of the universe $r
\sim 10^{27}cms$, shows an acceleration of $\sim 10^{-7}cm/sec$
which should be there everywhere, as indeed we are now coming to
learn (Cf.ref.\cite{smolin}).
\section{The Cosmological Constant Problem}
Let us now come to the cosmological constant problem. In the
author's 1997 cosmology referred to, we get a small cosmological
constant which is of the order
\begin{equation}
(\Lambda) \sim + 10^{-50} m^{-2}\label{164}
\end{equation}
This prediction was confirmed by the observations of Perlmutter and
other groups in 1998. The problem is that the known scales in
physics give a completely different value viz.,
\begin{equation}
Planck \, Scale \, (K^2 = \frac{c^3}{G\hbar}) \Rightarrow \Delta
\Lambda = \frac{c^3}{G\hbar} \sim 10^{121} \times 10^{-50}
m^{-2},\label{165}
\end{equation}
\begin{equation}
Z Boson \, Mass \, (K = \frac{mzc}{\hbar}) \Rightarrow \Delta
\Lambda = \frac{Gcm^4_z}{\hbar^3} \sim 10^{53} \times 10^{-50}
m^{-2},\label{166}
\end{equation}
\begin{equation}
Electron \, Mass \, (K = \frac{m_cc}{\hbar}) \Rightarrow \Delta
\Lambda = \frac{Gcm^4_c}{\hbar^3} \sim 10^{32} \times 10^{-50}
m^{-2}.\label{167}
\end{equation}
We note that (\ref{165}) is the same as \cite{aldro},
\begin{equation}
\Lambda_P = \frac{3}{l^2_P},\label{e15}
\end{equation}
In fact using the fact that $\Lambda$ depends on the energy density,
let us define the Planck energy density \cite{aldro}
\begin{equation}
\epsilon_P = \frac{m_P c^2}{(4 \pi / 3)l^3_P},\label{e16}
\end{equation}
where $m_P$ is the Planck mass. In terms of $\epsilon_P$ we have
from (\ref{e15}) and (\ref{e16})
\begin{equation}
\Lambda_P = \frac{4\pi G}{c^4} \epsilon_P\label{e17}
\end{equation}
Equation (\ref{e17}) can be considered to be the extreme case of a
local cosmological constant at the Planck scale. We can consider on
the contrary the cosmological constant,
\begin{equation}
\Lambda = \frac{4 \pi G}{c^4} \epsilon\label{e18}
\end{equation}
where in (\ref{e18}) we take for $\epsilon$ the average density of
the universe
\begin{equation}
\epsilon = Mc^2 \left[\left(\frac{4 \pi}{3}\right) R^3\right]^{-1}
(M \sim 10^{55}gm;\, R \sim 10^{27}cm).\label{26}
\end{equation}
Using (\ref{26}) in the above, it is easy to verify that we get the
correct value of the cosmological constant (Cf.ref.\cite{brax}). In
this connection we
also note the following:\\
A few years ago the author pointed out \cite{bgscosmic} that this
long standing puzzle can be resolved if we consider the cosmic
neutrino background as primary. In fact there has been mounting
evidence for such a cosmic background of neutrinos \cite{weiler}. In
fact earlier the author had shown that many neutrino parameters
including its mass could be obtained on the basis of fluctuations in
such a cold neutrino background \cite{fpl1,hayakawa}. It is believed
that the $GZK$ photo pion process seems to be the contributing
factor. With this background, let us now consider this neutrino
background to deduce the correct cosmological constant. We note that
the cosmological constant is given by
\begin{equation}
\lambda = < 0|H|0> \equiv \mbox{cosmological \, constant}\label{ey}
\end{equation}
The cosmological constant $\lambda$ is now given by its familiar
expression \cite{weinbergqtl}
\begin{equation}
\Lambda = \int^\lambda_0 \frac{4 \pi p^2}{(2\pi)^3} dp \frac{1}{2}
\sqrt{p^2 + m^2}\label{xe2}
\end{equation}
In (\ref{xe2}) $\lambda$ is the cut off which takes care of the
divergent integral. If we now use the value of the neutrino mass
$\sim 10^{-3}eV$ in (\ref{xe2}) then we get the value of the
cosmological constant as
\begin{equation}
\Lambda \sim 10^{-50}GeV^4\label{xe3}
\end{equation}
which is consistent with the latest observations pertaining to the accelerating universe with a small cosmological constant.\\
On the other hand, in the usual theory, $\lambda$ has been taken to
correspond to the Planck scale and the Planck mass $\sim
10^{19}GeV$. This has lead to the value of the cosmological constant
which is $10^{120}$ times its actual value as can be given in
(\ref{165}). This is the famous cosmological constant problem
\cite{weinbergphysrev}.\\
We can now see that by considering the cosmic neutrino background
rather than the Planck cut off, we get the right order of the
cosmological constant. This is related to the above approach
because, it is known that there are $\sim 10^{90}$ neutrinos in the
universe with a mass $\sim$ mass of the universe, given the modern
estimate of the neutrino mass. So the average density of the
universe, using the neutrino content, comes out to be the same as in
(\ref{26}).\\
Further references to the cosmological constant may be found in
\cite{goldsmith,perl,carroll,perl2} (and references therein).

\end{document}